# Experimental study of heterogeneous organic chemistry induced by far ultraviolet light: Implications for growth of organic aerosols by $CH_3$ addition in the atmospheres of Titan and early Earth

Peng K. Hong[a, b], Yasuhito Sekine[c], Tsutoni Sasamori[d], Seiji Sugita[c, d, e]

[a] Planetary Exploration Research Center, Chiba Institute of Technology, 2-17-1 Tsudanuma, Narashino, Chiba 275-0016, Japan

[b] Department of Systems Innovation, The University of Tokyo, 7-3-1 Hongo, Bunkyo, Tokyo 113-8656, Japan

[c] Department of Earth and Planetary Science, The University of Tokyo, 7-3-1 Hongo, Bunkyo, Tokyo 113-0033, Japan

[d] Department of Complexity Science and Engineering, The University of Tokyo, 5-1-5 Kashiwanoha, Kashiwa, Chiba 277-8561, Japan

[e] Research Center for the Early Universe, The University of Tokyo, 7-3-1 Hongo, Bunkyo, Tokyo 113-0033, Japan

Abstract

Formation of organic aerosols driven by photochemical reactions has been observed and suggested in $CH_4$-containing atmospheres, including Titan and early Earth. However the detailed production and growth mechanisms of organic aerosols driven by solar far ultraviolet (FUV) light remain poorly constrained. We conducted laboratory experiments simulating photochemical reactions in a $CH_4$-$CO_2$ atmosphere driven by the FUV radiations dominated by the Lyman-α line. In the experiments, we analyzed time variations in thickness and infrared spectra of solid organic film formed on an optical window in a reaction cell. Gas species formed by FUV irradiation were also analyzed and compared with photochemical model calculations. Our experimental results show that the growth rate of the organic film decreases as the $CH_4/CO_2$ ratio of reactant gas mixture decreases, and that the decrease becomes very steep for $CH_4/CO_2 < 1$. Comparison with photochemical model calculations suggests that polymerizations of gas-phase hydrocarbons, such as polyynes and aromatics, cannot account for the growth rate of the organic film but that the addition reaction of $CH_3$ radicals onto the organic film with the reaction probability around $10^{-2}$ can explain the growth rate. At $CH_4/CO_2 < 1$, etching by O atom formed by $CO_2$ photolysis would reduce or inhibit the growth of the organic film. Our results suggest that organic aerosols would grow through $CH_3$ addition onto the surface during the precipitation of aerosol particles in the middle atmosphere of Titan and early Earth. On Titan, effective $CH_3$ addition would reduce $C_2H_6$ production in the atmosphere. On early Earth, growth of aerosol particles would be less efficient than those on Titan, possibly resulting in small-sized monomers and influencing UV shielding.

# 1. Introduction

Photochemically-produced organic aerosols are found in the $CH_4$-containing atmospheres of Titan and giant planets in the outer Solar System (e.g., Gautier and Owen, 1989; Tomasko and West, 2009). Organic aerosols may have been also produced in $CH_4$-$CO_2$ atmospheres on early Earth and exoplanets (e.g., Sagan and Chyba, 1997; Trainer et al., 2006; Wolf and Toon, 2010; Kreidberg et al., 2014; Knutson et al., 2014.). Given the roles of these aerosols on climate, atmospheric structure, and compositions (e.g., McKay et al., 1989; Pavlov et al., 2001; Sekine et al., 2008a, 2008b; Lavvas et al., 2008a, 2008b; Wolf and Toon, 2010), it is important to understand how the atmospheric composition and energy source affect the production and growth of organic aerosols in planetary atmospheres.

In the present study, we focus on far ultraviolet (FUV) light (i.e., 120–200 nm in wavelength) as an energy source for the production and growth of organic aerosols in a $CH_4$-containing atmosphere. This is because the FUV light, predominated by the Lyman-α emission at 121.6 nm, is considered as the major flux dose that dissociates $CH_4$ and $CO_2$ in planetary atmospheres (e.g., Kasting et al., 1979; Trainer et al., 2006). On Titan, for instance, the solar FUV radiation is the dominant energy source in the middle atmosphere at 500–800 km in altitude, where direct photodissociation of $CH_4$ takes place (Krasnopolsky, 2009). At these altitudes, it has been suggested that the formation and growth of aerosol monomers would take place through gas-phase reactions as well as through surface reactions with radicals and neutral gas species (e.g., Lavvas et al., 2008b; 2011).

Previous laboratory experiments have simulated hydrocarbon photochemistry and organic aerosol formation driven by the solar FUV radiation using a deuterium lamp (Ádámkovics and Boering, 2003; Trainer et al., 2006; 2012; 2013; Hasenkopf et al., 2010). However, the spectrum of a typical deuterium lamp has distinctive strong emission lines around 160 nm due to $D_2$ molecular emissions, which is not significant in the solar FUV flux (Fig. 1a). In contrast, the Lyman-α line at 121.6 nm dominates the solar FUV radiation. Since $CH_4$ and $CO_2$ have distinct absorption cross sections around 160 nm (Fig. 1b), the use of a deuterium lamp may cause different photochemical reactions, compared with those by actual solar FUV. For instance, $CO_2$ has a significant absorption at 150–160 nm, while that of $CH_4$ is weak. This could result in a difference in $CH_4/CO_2$ dependency of organic aerosol production between in laboratory experiments using a deuterium lamp and in actual planetary atmospheres.

In previous studies, photochemical models also have employed to calculate the production rates of organic aerosols by solar FUV in $CH_4$-containing atmospheres (Lavvas et al., 2008a, 2008b; Krasnopolsky, 2009). To estimate the production rate of organic aerosols, these models assume some key polymerization reactions involving $C_4$–$C_6$ hydrocarbons as the main formation process for organic aerosols (Pavlov et al., 2001; Lavvas et al., 2008a, 2008b). However, the relationship between the rates of these reactions and aerosol particle formation remains poorly examined by laboratory experiments. Moreover, $C_1$–$C_6$ hydrocarbons are photolyzed by solar FUV, forming hydrocarbon radicals (e.g., Lavvas et al., 2008b; Krasnopolsky, 2009). These radicals could be consumed via heterogeneous reactions on the surface of organic aerosols and play a role for the growth of organic aerosols (e.g., Lavvas et al., 2008b). In photochemical models, nevertheless, these surface reactions are not taken into account (e.g., Krasnopolsky, 2009) or are calculated with hypothesized reaction rates (e.g., Lavvas et al., 2008b) due to the lack of experimental data.

In the present study, we report results of laboratory experiments on photochemistry and heterogeneous reactions of C-bearing radicals for organic solid production in $CH_4$-$CO_2$ gas mixtures by FUV radiation using a hydrogen-helium ($H_2$-He) FUV lamp, which has a UV spectrum dominated by the Lyman-α emission similar to the actual solar radiation (see Sec. 2.1.2 for details). We investigate the growth rate and infrared spectra of solid organic film formed on the optical window in the reaction cell for various $CH_4/CO_2$ ratios of the reactant gas mixtures. We also perform gas analysis and photochemical calculations to examine the mechanisms of the organic film growth in the experiments, especially focusing on heterogeneous reactions on the surface. We describe the experimental methods and photochemical model in Sec. 2 and present the results in Sec. 3. In Sec. 4, we discuss possible reactions of gas-phase chemistry that can produce the organic film, and the role of heterogeneous reactions on the organic film growth. We also provide implications of our results for stratospheric chemistry and organic aerosol production on Titan and early Earth in Sec. 4.

## 2. Methods

In this section we describe our experimental apparatus and methodology for organic photochemical reactions and analyses (Sec. 2.1). We also describe the one-box photochemical model for simulating the photochemical reactions of our experiments (Sec. 2.2).

### 2.1. Photochemical experiments

*2.1.1. Experimental configuration and procedure*

Figure 2 shows a schematic diagram of the experimental apparatus for photochemistry and production of solid organic compounds used in the present study. The system is an open-flow system, which consists of mainly two quartz glass tubes (Makuhari Rikagaku Glass Inc.) used for a reaction cell and FUV light source, respectively (Fig. 2). The reaction cell and FUV light source are separated by a $MgF_2$ window with 1-mm thickness (IR SYSTEM Co., Ltd.) having UV cutoff at 110 nm in wavelength. Thus, UV light at > 110 nm from the FUV light source can transmit into the reaction cell through the window, whereas gas species in each glass cell cannot be mixed.

Before the experiments, we evacuated both the reaction cell and FUV light source with two rotary pumps (Fig. 2). The reaction cell and FUV light source had a background pressure of $10^{-3}$ Torr (~$10^{-3}$ mbar) or less, as measured by thermocouple gauges (Varian Inc., ConvecTorr P-Type). Then, reactant gas mixtures were introduced into the reaction cell constantly through mass flow controllers (KOFLOC, Model 3200 series) from gas cylinders. The introduced reactant gas mixtures were evacuated with the rotary pump continuously. A $H_2$-He gas mixture was also introduced to the FUV light source through a mass flow controller, and evacuated with the other rotary pump. An optical fiber of an ultraviolet/visible (UV/VIS) spectrometer (Ocean Optics, USB 2000) was inserted at the opposite end of the reaction cell to the FUV light source to measure the time evolution of attenuation of FUV flux from the lamp. The UV/VIS spectrometer was also separated by a $MgF_2$ window from the reaction cell.

In the reaction cell, the reactant gas mixtures were photolyzed with FUV light, and radicals were produced by the FUV initiated chemical reactions. Given the optical depth of the Lyman-α emission in our reaction cell (typically, ~20 mm) and length of the reaction cell (260 mm), almost all of the FUV energy from the light source was absorbed within the reaction cell, especially in the vicinity of the FUV light source. By FUV irradiations, solid organic compounds were generated on the surface of the $MgF_2$ window which is inserted between the reaction cell and lamp, forming a thin organic film. We measured thickness and roughness of the films with a spectroscopic ellipsometer (HORIBA Jobin Yvon, Auto SE Lambda 650) (Sec. 2.1.3). A part of the reacted gas species in the reaction cell was introduced to a quadrupole mass spectrometer, or QMS (ULVAC, Qulee CGM), to perform mass spectrometry (Sec. 2.1.4) (Fig. 2). Chemical structures of the organic films deposited on the $MgF_2$ windows were analyzed with infrared spectroscopy (Sec. 2.1.5). All of the experiments were performed at room temperature of around 20ºC.

*2.1.2. Far ultraviolet light source*

In order to simulate the solar FUV light, we added noble gases into $H_2$ gas and excited the gas mixtures with a power source. Since high-energy meta-stable noble gases can collide with $H_2$ and dissociate it effectively into atomic hydrogen, addition of noble gases can suppress molecular emission lines of $H_2$ (Davis and Braun, 1968; Boduch et al., 1992). We used a premixed $H_2$-He gas mixture with $H_2$ : He = 1 : 9 (Purity 99.9999%, Japan Fine Products Co.). The gas mixture was introduced into the FUV light source at a pressure of 1 Torr, or 1.3 mbar. The $H_2$-He gas mixture was excited by a radio frequency (RF) power source (13.56 MHz, Nihon Koshuha Co., Ltd.). Previous studies show that the use of He gas results in a strong Lyman-α emission line compared with other noble gases, such as Ar, due to the high energies of meta-stable He atoms (Kuroiwa et al., 1992;

Rahman et al., 2004). Kuroiwa et al. (1992) measured the emission spectrum of a $H_2$-He lamp with $H_2 : He = 1 : 20$, which has a strong Lyman-α emission line similar to the actual solar UV flux (Fig. 1a). With the similar experimental settings with Kuroiwa et al. (1992), we considered that the Lyman-α line would also dominate the ultraviolet light irradiated from our FUV light source. This was confirmed by $N_2O$ and $CO_2$ actinometry, as described below.

In order to obtain the actinic flux, we measured the energy flux emitted from the FUV light source by performing $N_2O$ and $CO_2$ actinometry (e.g., Rajappan et al., 2010). The actinic flux was obtained from the photodissociated amounts of $N_2O$ or $CO_2$ measured with QMS, together with the cross sections and quantum yields of those gases obtained previously (Nakata et al., 1965; Okabe, 1967). The $N_2O$ gas is more selective to the Lyman-α emission than $CO_2$ since the absorption cross section of $N_2O$ around 160 nm is about one order of magnitude smaller than $CO_2$ (Fig. 1b). The overall quantum yield of $N_2O$ photolysis is estimated as 1.5 $N_2O$-molecule per photon considering the decomposition by both photolysis and subsequent reactions (Rajappan et al., 2010). The photolysis of a $CO_2$ molecule results in the production of a CO and O atom. Since the reaction of O atom with $CO_2$ or CO molecule is extremely slow (Tsang and Hampson, 1986), the photolysis of $CO_2$ is the primary loss reaction for $CO_2$. Thus, the overall quantum yield for $CO_2$ actinometry was assumed to be one $CO_2$-molecule per photon.

We performed the $N_2O$ and $CO_2$ actinometry with the RF power of our FUV light source ranging from 20 W to 120 W in order to obtain the relationship between the RF power input and resulting FUV flux. The pressures of $N_2O$ or $CO_2$ gas in the reaction cell ranged 1.0–6.5 Torr, or 1.3–8.5 mbar. Pure $N_2O$ gas (purity 99.9%, Suzuki Shokan Co., Ltd.) or $CO_2$ gas (purity 99.999%, Taiyo Nippon Sanso Corporation) was introduced into the reaction cell and irradiated with the FUV light source. The QMS signal at $m/z = 44$ for $N_2O$ or $CO_2$ gas was measured before and during the UV irradiations to derive the photodissociated amounts of the reactant gas. A typical example of QMS signal during the UV irradiation is shown in the supplementary material.

Figure 3 shows the relationship between the RF power input and actinic flux generated by the FUV light source, measured with $N_2O$ and $CO_2$ actinometry. The magnitude of the actinic flux ranges in the order of $10^{13}$–$10^{14}$ photon cm$^{-2}$ s$^{-1}$, which is consistent with the results using a similar, plasma-induced UV source with $H_2$-He gas mixtures (Westley et al., 1995). This figure shows that the actinic flux increases with the RF power, but the gradient of the actinic flux dampens at >90 W. This may be caused due to self-absorption of the Lyman-α emission by H atoms in the ground state. According to Yamashita (1975), larger numbers of H atoms in the ground state are generated at higher input RF power via more effective $H_2$ dissociations.

Figure 3 also shows that $CO_2$-derived actinic flux is within the variance of $N_2O$-derived actinic flux, indicating that there is no significant difference in the actinic fluxes whether using $N_2O$ or $CO_2$. This indicates that the UV spectra of our FUV light source are dominated by the Lyman-α line, not by molecular emission lines around 160 nm, as the absorption cross section of $N_2O$ gas at around 160 nm is about one order of magnitude smaller than $CO_2$ (Fig. 1).

*2.1.3. Reaction cell and measurements of growth rate*

Mixtures of $CH_4$ (purity 99.999%, Japan Fine Products Corporation) and $CO_2$ were introduced into the reaction cell through the mass flow controllers with 1.5 mL min$^{-1}$ of total flow rate. We varied the $CH_4/CO_2$ ratio of the gas mixtures by changing their flow rates. The total pressure in the reaction cell was fixed to be 5 Torr, or 6.5 mbar. Although direct dissociation of $N_2$ occurs only by high-energy particles or EUV radiation, Trainer et al. (2012) and Yoon et al. (2014) report incorporation of N into organic aerosols through FUV radiation onto $N_2$-$CH_4$-containing gas mixtures, possibly through indirect dissociation of $N_2$. Thus, inclusion of $N_2$ gas could affect formation rate of organic aerosol in planetary atmospheres. However, the present study aims to examine heterogeneous reactions of C-bearing radicals and to investigate $CH_4$-$CO_2$ photochemistry in the middle atmospheres, by comparing results of laboratory experiments with photochemical model. Thus, we did not include $N_2$ gas in the gas mixture.

In order to obtain the growth rate of organic film formed on the $MgF_2$ window in the reaction cell, we measured its thickness with spectroscopic ellipsometry. After the FUV irradiation, the $MgF_2$ window was removed from the reaction cell. The organic film was deposited on the side of the $MgF_2$ window exposed to the reaction cell. The configuration of the ellipsometry measurements was as follows; the incident angle of the beam was $66° \pm 0.5°$, wavelength region of the beam was from 425 nm to 1000 nm, and the diameter of the beam was 1 mm. The beam was irradiated onto the center of the $MgF_2$ window set in the ellipsometer.

In ellipsometry, the amplitude ratio ($\Psi$) and phase difference ($\Delta$) upon reflection of the beam on the organic films were measured. In the data analyses, both thickness and optical property of the organic films were calculated to fit the measured $\Psi$ and $\Delta$. The fitting of the ellipsometry data was performed with a commercialized software Delta Psi2 (Horiba Scientific Co.). We employed a two-layer model for the data analysis of the ellipsometry, in which the produced organic film was composed of an upper thin, partially oxidized hydrocarbon layer and a thick lower un-oxidized hydrocarbon layer. This was the case because the surface of an organic film could have been oxidized when it was exposed to the air (Sciamma-O'Brien et al., 2010). However, since a single-layer model without an upper oxidized layer calculated the thickness of organic film close to those obtained by the two-layer model within 1% deviation, the influence of oxidation on film thickness was not significant in our experiments.

We conducted two distinct methods for obtaining the growth rate of organic film. In Method 1, we measured time variations in thickness at arbitrary time intervals during the experiments, typically a few minutes. With this method, we obtained a film thickness directly as a function of time; however, the sample was exposed to the air at each measurement. To avoid the air exposure, in Method 2, we measured time variations in UV flux at 300 nm from the FUV light source using the UV/VIS spectrometer installed at the end of the reaction cell (Fig. 2). Upon the growth of organic film during the experiment, the UV flux at 300 nm decreased due to UV shielding by the growing film. The film thickness was measured only after the end of experiment. With this method, the samples were not exposed to the air during the experiments. With the measured thickness after the experiments and time-varied UV flux from the FUV light source, we indirectly calculated time evolution of film thickness assuming a constant absorption coefficient of the film at 300 nm. After a few hours of irradiation, film growth becomes slow due to UV shielding by the growing film (See Supplementary Material). We thus used the experimental data before two hours of irradiation time.

### 2.1.4. Mass spectroscopy of gas-phase products

We measured the abundances of gas species after the FUV irradiation in order to investigate hydrocarbon photochemistry for various $CO_2/CH_4$ gas mixtures. The reacted gas mixtures were introduced into the QMS through a variable leak valve (Fig. 2). The chemical compositions of the reacted gas mixtures reached a steady state typically after one minute of reaction time. The mass spectra were obtained after reaching the steady states.

We performed mass deconvolution of the obtained mass spectra (e.g., Ádámkovics and Boering, 2003). The gas species considered in the deconvolution analyses are $H_2$, $CH_4$, $C_2H_2$, $C_2H_4$, $C_2H_6$, $C_3H_4$, $C_3H_6$, $C_3H_8$, $C_4H_2$, $C_4H_{10}$, $C_6H_6$, CO, $CO_2$, $H_2O$, $H_2CO$, $CH_3OH$, $CH_3CHO$, $O_2$ and $N_2$. The fragmentation patterns of $CH_4$, $C_2H_2$, $C_2H_4$, $C_2H_6$, CO, and $CO_2$ were obtained by introducing pre-mixed standard gas mixtures with Ar gas into the QMS. To obtain the quantitative mixing ratios of these species in a reacted gas mixture (i.e., calibration), we changed the amount of standard gas introduced into the QMS and obtained the relationships between the partial pressure and intensity of the mass spectra for these species. The fragmentation patterns of the other gas species were obtained from the database of National Institute of Standards and Technology[1]. Deconvolution analyses were based on those given by Ádámkovics and Boering (2003) and were conducted by manually fitting the observed spectra with linear mixings of the species considered (also see Supplementary Materials).

## 2.1.5. Infrared spectroscopy of organic films

In order to analyze the chemical bonds incorporated in the organic films formed on the $MgF_2$ windows, we measured their infrared transmittance spectra with a Fourier transform infrared, or FTIR, spectrometer (PerkinElmer, Inc., Frontier FT-NIR/MIR) in the ambient laboratory atmosphere. Infrared spectra were obtained in the range of 4000 to 450 cm$^{-1}$ in the wave number with a resolution of 1 cm$^{-1}$, and 5-minute scans were averaged. A MgF$_2$ substrate without a sample was used as a reference. Absorptions due to water vapor and CO$_2$ gas in the ambient atmosphere were calibrated with a commercialized software Atmospheric Vapor Compensation$^{TM}$ (PerkinElmer, Inc.).

---

[1] http://webbook.nist.gov/chemistry/

*2.2. Photochemical model*

We developed a one-box photochemical model that includes 134 species up to C$_8$ hydrocarbons and 791 chemical reactions (see Appendix for the list of reactions). Upon compiling the reaction list, we employed the rate constants of hydrocarbon chemistry used in the atmosphere of Titan (Hébrard et al., 2006). We also used the reaction rates of O-bearing species, such as alcohols and aldehydes, based on a photochemical model about early Earth (Pavlov et al., 2001) in which some of the rate constants are updated (Burkholder et al., 2015). The photochemical scheme contains only neutral chemistry. It also includes the key polymerization reactions for formation of aerosol monomer assumed in the previous studies (Yung et al., 1984; Toublanc et al., 1995; Pavlov et al., 2001; Wilson and Atreya, 2004; Hébrard et al., 2006; Lavvas et al., 2008a, 2008b; Krasnopolsky, 2009, 2010), which are summarized in Table 1. Some of the polymerization reactions have been assumed based on combustion chemistry of hydrocarbons (Richter and Howard, 2000).

To compare our experimental results of photochemically-produced gas species in the reaction cell, we considered a one-box open system with continuous inflow and outflow, similar to the experimental system. The outflow rate $u$ was estimated to be 0.11 s$^{-1}$ with the inflow rate (1.5 mL min$^{-1}$) and the total volume of the reaction cell (3.3 × $10^2$ mL), based on the experimental setup. With the estimated outflow rate $u$ s$^{-1}$, the outflow flux $f$ molecule cm$^{-3}$ s$^{-1}$ can be expressed as

$$f = u \cdot n,$$

where $n$ is the number density in molecule cm$^{-3}$. We fixed the total pressure of the box to be 5 Torr, or ~6.5 mbar, as same as our experimental condition. We assume no UV absorption by organic film in the photochemical calculations. This is the case because the QMS spectra is obtained in a few minutes after the initiation of UV irradiation, and because organic film is too thin to absorb UV light significantly within this reaction time. Our results of time evolution of QMS show that their signal intensities reach steady states in a few minutes after an initiation of UV irradiation (Figure S5). The time evolution of the mass spectra shows that the gas phase composition has been stable throughout the experiments (Figs. S4 and S5 in Supplementary Materials). In later stage of the experiment, e.g., after a few hours since the irradiation, film growth becomes slow especially for low $CH_4/CO_2$ experiments. We thus model the experiments only for the early stage of experiments when the growing film thickness does not significantly affect the transmission of UV light. We assume that the UV spectrum in the calculations is same as solar UV light given the similarity of the UV spectra (Fig. 1). Sensitivity of the UV spectrum to the results is discussed in Supplementary Materials (section 3). The photochemical calculation was initiated with a time step of $10^{-7}$ sec, which was automatically adjusted after each calculation. Although the calculation was carried out at least 10 hours after the irradiation begins, steady states were achieved after about a minute after the irradiation, similar to the gas compositions measured with QMS (see Sec. 2.1.4).

## 3. Results

*3.1. Dependence of organic film growth rate on $CH_4/CO_2$ ratio*

Figure 4 shows the film thickness as a function of irradiation time for the experiment using 100% $CH_4$ gas with RW power of 90 W. Since the thickness increases linearly with irradiation time, we employed the slope of the best-fit linear function as the growth rate of organic film. The film thickness was obtained by Methods 1 and 2 described in Sec. 2.1.3. In the ellipsometry analysis, the bet-fit of the measured $\Psi$ and $\Delta$ is obtained when considering no porosity in the upper layer of an organic film. This suggests that the organic films formed in the present study would have a microscopically flat surface, rather than an accumulation of aerosol particles (Fujiwara et al., 2001).

Figure 5 shows that the growth rates of the organic films from $CH_4$-$CO_2$ gas mixtures as a function of $CH_4/CO_2$ ratio of the reactant gas mixtures. The RF power for these experiments was fixed at 90 W, corresponding to the photon flux of $1.8 \times 10^{14}$ photon cm$^{-2}$ s$^{-1}$ in the reaction cell (Fig. 3). Despite the difference in the measurement methods of organic film thickness, namely Methods 1 and 2 (see Sec. 2.1.3), the obtained results are consistent with each other, indicating that the oxidization of the surface by air does not affect the thickness of organic films significantly.

Figure 5 shows that the growth rate decreases with decreasing $CH_4/CO_2$, showing a gradual decrease from pure $CH_4$ to $CH_4/CO_2 > 3$ and a steep decrease where $CH_4/CO_2 < 2$. The observed relationship between the growth rate of the organic film and $CH_4/CO_2$ ratio is consistent with the previous experiments on organic aerosol formation from $CO_2$-$CH_4$ gas mixtures using an electric discharge (Trainer et al., 2004). However, our results are different from those obtained by Trainer et al. (2006), in which organic aerosols are produced by FUV irradiations with a deuterium lamp. Trainer et al. (2006) report that the aerosol production rate has an optimal value at around the $CH_4/CO_2$ ratio of unity (Fig. 5). The discrepancy between the present study and Trainer et al. (2006) may be caused by the differences in the experimental conditions. The primary difference in the experimental conditions between the present study and Trainer et al. (2006) is the spectra of UV light source; namely, the present study used a $H_2$-He lamp, whereas Trainer et al. (2006) used a deuterium lamp. Compared with a deuterium lamp, FUV light from a $H_2$-He lamp dissociates $CH_4$ more effectively than $CO_2$ given the absorption cross section of these gas species (Fig. 1). This would lead to efficient organic solid production for the experiments using a $H_2$-He lamp even at lower $CH_4/CO_2$ ratios. Nevertheless, this is not the case as shown in Fig. 5. Thus, the difference in the UV light may not simply explain the discrepancy.

One possible explanation for the difference between the present study and Trainer et al. (2006) is the effect of total pressure. We performed the experiments at total pressure of 5 Torr; whereas, Trainer et al. (2006) performed at 600 Torr. The rate coefficients of three-body reactions become greater at higher pressures, which leads to more efficient production of saturated hydrocarbons and, thereby, less aerosol formation (e.g., Imanaka et al., 2004). Under high-pressure conditions of Trainer et al. (2006), H removal from saturated hydrocarbons by O atoms formed by $CO_2$ photolysis would have promoted aerosol production even at relatively high $CO_2$ concentrations. The difference of total pressure also results in a different residence time for gas species inside the reaction cell. The residence time of our experiments is ~10 s estimated with the inverse of outflow rate $u$, while the residence time of Trainer's experiments is ~300 s (Trainer et al., 2006; Hasenkopf et al., 2010). Thus, the shorter residence time of our experiments may have caused a production yield lower than Trainer's experiments.

Another possible explanation is the difference in the formation mechanism of organic solids. Trainer et al. (2006) obtained the aerosol production rate based on weight measurements of aerosol particles. Thus, the measured production rate by Trainer et al. (2006) may be also affected by a nucleation process of aerosol particles in the gas phase. On the other hand, the present study measures the growth rate of an organic film formed on the $MgF_2$ window. Accordingly, the growth rate obtained by the present study could be largely controlled by surface reactions between the solids and gas molecules. The chemical networks and parent molecules may be different between the nucleation and surface growth, possibly leading to difference in the dependence on $CH_4/CO_2$ ratio. In fact, in our experimental configuration (relatively short residence time of gas: ~10 s) and sample collection (deposits on a $MgF_2$ window), the organic films formed in the present study might tend to be generated largely by surface heterogeneous reactions. More detailed discussion on the role of heterogeneous reactions in the present study is shown below in Sec. 4.1.2.

An inclusion of $N_2$ gas into initial gas mixtures could have caused the difference in the formation rates between Trainer et al. (2006) and our experiments. In fact, Trainer et al. (2012) and Yoon et al. (2014) reported nitrogen incorporation into aerosol particles by FUV irradiation. They suggest that reactions of $N_2$ with CH radicals may have caused the production of CN radicals in the gas phase, which can polymerize into high-molecular-weight hydrocarbons. If formation of CN radicals occurs effectively at around $CH_4/CO_2$ ~1, this could have resulted in high production rate in the experiments by Trainer et al. (2006).

*3.2. Composition of gas-phase products*

The results of mass spectra of the gas-phase products by the Lyman-α irradiations onto various $CH_4/CO_2$ gas mixtures are shown in Fig. 6. At $CH_4/CO_2 > 1$, relatively high-molecular-weight hydrocarbons, such as $C_6$ and $C_7$ hydrocarbons, are produced as shown in the mass peaks at *m/z* 78–87 and 91–100. The mass peaks at *m/z* 78 and 91 are consistent with signals of $C_6H_6^+$ (phenyl derivative) and $C_7H_7^+$ (benzyl derivative). In contrast, these relatively high-molecular-weight hydrocarbons are not generated at $CH_4/CO_2 < 1$, indicating that hydrocarbon formation is suppressed possibly by O-bearing species, including $CO_2$ and its photolysis products. Other notable signals in the mass spectra include peaks at *m/z* 15 ($CH_3^+$), 16 ($CH_4^+$, $O^+$), 26 ($C_2H_2^+$), 28 ($C_2H_4^+$, $CO^+$, $N_2^+$), 30 ($C_2H_6^+$), 39 ($C_3H_3^+$), 41 ($C_3H_5^+$), 43 ($C_3H_7^+$, $C_2H_3O^+$) and 44 ($CO_2^+$).

Figure 7 shows the molar mixing ratios of major gas species retrieved with the deconvolution analysis of the mass spectra shown in Fig. 6. Only the abundance of $C_6H_6$ was estimated by QMS intensity at m/z = 78 and normalized to pure $CH_4$ condition. This figure shows that the mixing ratios of $C_2$ hydrocarbons are almost constant at $CH_4/CO_2 \geq 2$ and decrease when the $CH_4/CO_2$ ratio becomes less than unity. In contrast, the mixing ratio of CO increases with decline in $CH_4/CO_2$ ratio, as CO is a photochemical product of $CO_2$ photolysis. We also compare the results of the gas-phase products with those obtained by our photochemical calculations. Figure 7 shows that the results of photochemical calculations are consistent with the experimental results within the errors, suggesting that as far as up to $C_2$ species including O-bearing species, our photochemical model can reproduce the photochemical reactions occurred in the reaction cell with uncertainty within a factor of 2–3. The abundances of $CH_4$ are underestimated in the calculations because the UV spectrum of the light source employed in our experiments would emit relatively low flux of the Lyman-α line compared with the solar UV light, which is assumed as the light source in the calculations. The derived abundances of $C_6H_6$ are significantly lower than photochemical calculations for low $CH_4/CO_2$ conditions, suggesting that either $C_6H_6$ are quickly lost to form larger hydrocarbons or low-order hydrocarbons are quickly lost before forming $C_6H_6$. More detailed sensitivity study on the light source is described in Supplementary Materials. Our results show that the molar mixing ratios of $C_2H_2$, $C_2H_4$ and $C_2H_6$ decrease only about one order of magnitude when the $CH_4/CO_2$ ratio decreases from 10 to 0.4 (Fig. 7), whereas the observed growth rate of organic films decreases about two orders of magnitude from $CH_4/CO_2$ of 10 to 0.4 (Fig. 5).

*3.3. Infrared spectroscopy of organic aerosols*

Figure 8 shows typical infrared spectra of the organic films deposited on the $MgF_2$ windows. The spectra show characteristic absorption bands attributed to O–H stretching (3200–3600 cm$^{-1}$), $CH_3$ and $CH_2$ stretching (2800–3000 cm$^{-1}$), C=O stretching (1700–1800 cm$^{-1}$), C=C stretching (1620–1680 cm$^{-1}$), C–H deformation ($CH_2$; 1430–1470 cm$^{-1}$), and C–H deformation ($CH_3$; 1370–1390 cm$^{-1}$) (e.g., Hesse et al., 2007). The presence of O–H and C=O bonds suggests incorporations of O atoms in the organic films from the $CH_4/CO_2$ gas mixtures. Despite the presence of aromatic organic molecules, such as benzene, in the gas phase (Fig. 6), characteristic absorptions due to aromatic hydrocarbons at 3100–3000 cm$^{-1}$ (C-H stretching bands) and 1580–1560 cm$^{-1}$ (C=C stretching bands) (Hesse et al., 2007) are insignificant in our samples. The infrared spectra of the organic films for low $CH_4/CO_2$ ratios show very weak absorptions because the film thicknesses were too thin (tens of nanometers) for infrared spectroscopy. These results suggest that the organic film formed in the present study is likely to consist of a polymer-like olefinic structure of carbon and oxygen terminated with –CH$_3$ or –OH, rather than polyaromatic hydrocarbons.

The observed infrared spectra of the organic films formed by FUV irradiation (Fig. 8) are distinct from typical infrared spectra of the Titan aerosol analogues formed from N$_2$-CH$_4$ gas mixtures at low pressures (< 1 Torr) by cold plasma irradiations (e.g., Khare et al., 1984; Imanaka et al., 2004). These Titan aerosol analogues exhibit strong absorptions due to aromatic C=C and C≡N bonds at 1580–1560 cm$^{-1}$ (e.g., Imanaka et al., 2004). These organic aerosol analogues also contain fewer CH$_3$ and CH$_2$ bonds in their structures (Imanaka et al., 2004). On the other hand, the infrared spectra shown in Fig. 8 are similar to those of the Titan aerosol analogues formed at high pressures (> 1 Torr) by cold plasma, except for the lack of absorptions due to N-containing bonds in the Titan aerosol analogues (Imanaka et al., 2004).

**4. Discussion**

In this section, we first discuss the reaction mechanisms in the gas-phase toward the production of the organic films in our experiments. Then we discuss the role of heterogeneous reactions on the surfaces for the growth of the organic films. Finally we apply our experimental results photochemistry and organic aerosol production in the middle atmospheres of Titan and early Earth.

*4.1. Mechanism for formation of organic film*

*4.1.1. Parent gas molecules and polymerization reactions for formation of organic film*

As described above in Sec. 1, photochemical models have calculated the production rate of organic aerosols in $CH_4$-containing atmospheres, based on the conversion rates of $C_4$–$C_6$ hydrocarbons into higher-molecular-weight hydrocarbons (e.g., Lavvas et al., 2008b). These photochemical models propose that aromatic hydrocarbons and polyynes would be key parent molecules that are effectively polymerized to organic aerosols (Wilson and Atreya, 2004; Hébrard et al., 2006; Lavvas et al., 2008a, 2008b; Krasnopolsky, 2009, 2010). In the present study, we show that our photochemical model can reproduce the gas-phase chemistry in the reaction cell, at least the reactions involving $C_2$ hydrocarbons. Thus, based on comparison between the rates of polymerization reactions calculated with our photochemical model and the growth rate of the organic films obtained in the experiments, we discuss whether the reactions of proposed parent molecules can account for the formation of the organic films in our experiments.

Figure 9 shows the calculation results of the rates of polymerization reactions of alkene, aromatic hydrocarbons, and polyynes as a function of the $CH_4/CO_2$ ratio, compared with the normalized growth rate of the organic films. The dependency of the rates of polymerization reactions involving polyynes ($C_4H + C_8H_2$: R469 see Appendix for the reaction number, $C_6H + C_4H_2$: R511, $C_6H + C_6H_2$: R512, and $C_6H + C_8H_2$: R513) and O-bearing molecules ($O(^3P) + C_3H_3$: R661, $O(^3P) + C_3H_5$: R662, $CH_2OH + C_2H_4$: R783, and $CH_3CO + C_2H_3$: R814) does not match with that of the growth rate of the organic films (Fig. 9a and 9b). These reaction rates of polyynes and O-bearing molecules exhibit an optimal rate at around the $CH_4/CO_2$ ratio of unity (Fig. 9a). This is because polyynes are produced mainly from $C_2H$ that is partly produced through destruction of $C_2$ hydrocarbons by O and OH radicals. In addition, O-bearing molecules, such as $CH_3CO$ and $CH_2OH$, and $O(^3P)$ are also produced through the chemical networks initiated by $CO_2$ photolysis. Thus, the mixing ratios of these molecules and their reaction rates become optimal at $CH_4/CO_2 \sim 1$.

Among the proposed polymerization reactions, the observed steep decrease of the growth rate for $CH_4/CO_2 < 1$ is consistent only with the rates of the reactions involving aromatic hydrocarbons, such as $C_6H_6$ and $C_6H_5$, i.e., $C_2 + C_6H_6$: R310, $C_6H_5 + C_2H_2$: R515, $C_6H_5 + C_2H_2 + M$: R516, and $C_6H_5 + C_6H_6$: R517 (Fig. 9b). The polymerization reactions from $C_6H_6$ and $C_6H_5$ exhibit similar steep decreases with $CH_4/CO_2$ for the ratio < 1 (Fig. 9b). In our model, $C_6H_6$ and $C_6H_5$ are produced mainly by polymerization of $C_2H_2$ through $C_4H_5$ (also see Wilson et al., 2003; Krasnopolsky, 2014; Sciamma-O'Brien et al., 2014). Thus, a decline in the mixing ratio of $C_2H_2$ with $CH_4/CO_2$ is amplified for those of $C_6H_6$ and $C_6H_5$. This would lead to steep decreases in the polymerization reaction rates involving $C_6H_6$ and $C_6H_5$. The role of benzene as a potential precursor for Titan aerosol analogues has been recently reported by Trainer et al. (2013), Yoon et al. (2014), Sebree et al. (2014) and Sciamma-O'Brien et al. (2014).

Although the reaction rates involving $C_6H_6$ and $C_6H_5$ show a dependency on $CH_4/CO_2$ ratio similar to that of the growth rate of the organic films, they solely cannot account for the observed growth of the organic films. This is because the amounts of organic aerosols formed via these reactions are not quantitatively sufficient to account for the observed thickness of the organic films. For example, the growth rate of the organic film is measured as 18.5 nm min$^{-1}$ for pure $CH_4$ gas experiment with a RF power of 90 W (Fig. 5). On the other hand, the sum of the column production rates through the reactions of R515, R516, and R517 is calculated to be $8 \times 10^{12}$ cm$^{-2}$ s$^{-1}$, corresponding to a mass production rate of $1 \times 10^{-9}$ g cm$^{-2}$ s$^{-1}$. Assuming $\sim 1$ g cm$^{-3}$ of the density, which is approximately the average between the measured material density and particle density (Trainer et al., 2006; Imanaka et al., 2012; Hörst & Tolbert, 2013), the total growth rate of the organic film via the polymerization of aromatic hydrocarbons reaches to $\sim 8 \times 10^{-4}$ nm min$^{-1}$, which is $< 10^{-4}$ times the measured growth rate. The fact that the polymerization reactions of aromatic hydrocarbons are not the major mechanism for the growth of organic film is more compellingly supported by the lack of infrared absorptions due to aromatic hydrocarbons in the produced organic film (Fig. 8). The reaction rates of $C_2H + C_2H_2$ (R219) and $CH + C_4H_8$ (R314) are predominant among the polymerization reactions (Fig. 9c). Even considering the reactions of R219 and R314, the predicted growth rate reaches only $10^{-2}$–$10^{-3}$ times the measured value. Thus, with our experimental configuration, the growth of the organic films would not occur mainly through the polymerization reactions in the gas phase proposed by the previous models.

*4.1.2. Heterogeneous reactions on the surface*

In this subsection, we examine the role of heterogeneous reactions on the surface of organic films for its growth. Previous studies in applied physics have investigated the mechanisms and growth rates of amorphous hydrogenated carbon (a-C:H) films formed by cold plasma irradiations onto gaseous $CH_4$ (e.g., von Keudell & Jacob, 1996; von Keudell et al., 2000; 2001). These studies show that the reactions of both $CH_3$ radicals and H atoms with the surface cause the synergistic growth of a-C:H film (von Keudell et al., 2000; 2001); that is, a collision of H atom abstract hydrogen on the surface, creating a dangling bond where a $CH_3$ radical can add onto a-C:H film (von Keudell et al., 2001). Sekine et al. (2008a) show that H abstraction from Titan aerosol analogue by collisions of H atoms in the gas phase also occurs through the mechanism and reaction rate similar to those on a-C:H film. These previous studies imply that heterogeneous reactions of $CH_3$ radical and H atom with the organic film could have occurred for the growth of the film in our experiments.

To examine the role of heterogeneous reactions more quantitatively, we estimate the numbers of collision of $CH_3$ radicals onto the surface of organic film in our experiments. The kinetic theory of gases gives the collision flux of a gas molecule or radical with solid surface, $Z_W$ m$^{-2}$ s$^{-1}$, as follows,

$$Z_W = \frac{p}{\sqrt{2\pi m k_B T}} \quad (1)$$

where $p$ is the partial pressure of the gas or radical in Pa, $m$ is the atomic or molecular weight in kg, $k_B$ is Boltzmann constant, and $T$ is the temperature in K. The surface reaction rate, $R_W$ in m$^{-2}$ s$^{-1}$, is given by multiplying the collision flux with reaction probability $s$,

$$R_W = Z_W \cdot s. \quad (2)$$

von Keudell et al. (2001) have obtained the reaction probability for addition of a $CH_3$ radical with a-C:H film as ~$10^{-4}$ by irradiation of $CH_3$ radical alone. They also have shown that the reaction probability of $CH_3$ addition increases to ~$10^{-2}$ by simultaneous irradiations of $CH_3$ radical and H atom, due to the effect of hydrogen removal from the surface by H atoms (von Keudell et al., 2001). Thus, the reaction probability of $CH_3$ addition could range between $10^{-2}$ and $10^{-4}$ in our experiments.

Using the hypothesized reaction probability of $CH_3$ addition ($10^{-4}$–$10^{-2}$) and the number density of $CH_3$ radical in the reaction cell calculated by the photochemical model, the collision flux and addition rate of $CH_3$ radicals onto the surface of the organic film are calculated as ~$8 \times 10^{16}$ cm$^{-2}$ s$^{-1}$ and ~$8 \times 10^{12-14}$ cm$^{-2}$ s$^{-1}$, respectively, for the experiment using pure $CH_4$ gas. Assuming ~1 g/cm$^3$ for the density of the organic film, the growth rate due to $CH_3$ addition is estimated to be on the order of ~0.1–10 nm min$^{-1}$. Thus, the measured growth rate of the organic film (18.5 nm min$^{-1}$: Fig. 5) is consistent with the expected growth rate using the reaction probability of $CH_3$ addition of $10^{-2}$. The reaction probability of $10^{-2}$ is plausible in our experiments, as a large number of H atoms are also produced by $CH_4$ photolysis. The difference within a factor of 2 between the measured and expected growth rates may be caused by the assumption of our photochemical model, in which we consider the reaction cell as one box. The actual concentrations of $CH_3$ radical near the $MgF_2$ window of the reaction cell would be higher than the mean value in the reaction cell, as the $MgF_2$ window is set near the FUV source (Fig. 2). Based on the above discussion, we suggest that the growth of the organic films can be explained by the heterogeneous reaction of addition of $CH_3$ radicals for $CH_4$-rich gas mixtures (e.g., $CH_4/CO_2 > 3$). This is supported by the polymer-like olefinic structure of the organic film terminated with -$CH_3$, as revealed with the infrared spectroscopy (Fig. 8).

Figure 10 shows the calculation results of the reaction rate of $CH_3$ addition in the reaction cell of our experiments as functions of $CH_4/CO_2$ ratio, based on the number density of $CH_3$ radicals obtained by the photochemical model and assuming the reaction probability as $10^{-2}$. This figure shows that the rate of $CH_3$ addition does not exhibit a steep decrease for $CH_4/CO_2 < 1$, which is observed in the growth rate of the organic film. This implies the possible presence of competitive heterogeneous reactions that would reduce or inhibit the growth under $CO_2$-rich conditions. One possibility is that etching of the organic film by O atoms produced by $CO_2$ photolysis. In fact, O atoms are known to erode a-C:H film at high reaction probability (e.g., Bourdon et al., 1993). Bourdon et al. (1993) obtain the etching rate of ~20 ng cm$^{-2}$ s$^{-1}$ for a olefinic film for ~$10^{16}$ cm$^{-2}$ s$^{-1}$ of O atom flux, corresponding to ~$10^{-1}$ of the reaction probability of etching by O atoms. Figure 10 also shows the calculation results of the etching rate by O atoms, considering the number density of O atoms calculated with the photochemical model and assuming the reaction probability of $10^{-1}$. These results show that the rate of etching by O atoms becomes the same order of magnitude with that of $CH_3$ addition at around $CH_4/CO_2$ of ~1. These results may explain the steep decrease in the growth rate of the organic film at $CH_4/CO_2$ ~1. Although the rate of etching by O atoms exceeds that of $CH_3$ addition at $CH_4/CO_2$ less than ~0.3 (Fig. 10), we observed the formation of organic film even at $CH_4/CO_2$ ~0.3 or less. This may imply that at low $CH_4/CO_2$, O- and C-bearing radicals or reactive molecules, such as $CH_2OH$, $CH_3CO$, HCO, and formaldehyde, would become building materials for the organic film, as the produced organic film contains C=O and O-H bonds in their structures.

The FUV irradiation could have altered the chemical properties of the film by losing H-containing bonds, such as C-H, forming unsaturated bonds during the experiments. If this occurs effectively, continuous FUV irradiation could have formed dangling bonds on the surface of organic film, leading to direct addition of $CH_3$ by collision to the dangling bonds. However, our IR spectral measurements indicate that organic films after FUV irradiation exhibit the presence of C-H bonds, suggesting that the organic films are not fully altered by the FUV during the experiments (Fig. 8). In addition, we have produced organic films for irradiation times of 30 and 2 hours (Fig. S13). Given both the thickness of the organic films (2970 nm; reaction time of 30 hours and 550 nm; reaction time of 2 hours) and typical penetration depth of a gas molecule (a few nm: Sekine et al., 2011a), the inner parts of the films are largely isolated from the gas phase and continuously irradiated by FUV light during the experiments. Despite the difference in FUV irradiation time, Fig. S13 shows that the difference in FUV irradiation time does not significantly change the structure of these organic films, supporting that FUV irradiation does not affect the structure of organic film and suggesting our conclusion of $CH_3$ addition as the predominant growth process of organic film.

*4.2. Implications for $CH_4$-containing planetary atmosphere*

Our experimental results suggest that the solar FUV radiation onto a $CH_4$-containing atmosphere induce the growth of organic aerosols through heterogeneous reactions of $CH_3$ addition. The reaction probability of $CH_3$ radical with an organic aerosol analog would be $\sim 10^{-2}$. In the following section, we discuss the implications for these results for the formation and growth of organic aerosols in the atmospheres of Titan (Sec. 4.2.1) and early Earth (Sec. 4.2.2).

*4.2.1. Titan's atmosphere*

The observations by the Cassini spacecraft together and detailed photochemical models indicate that the formation of solid organic particles is initiated through ion and neutral photochemical reactions in the thermosphere (e.g., Waite et al., 2007; Krasnopolsky, 2009). The organic particles are considered to grow during the precipitation in the atmosphere through competing processes between surface growth, coagulation, and sedimentation (e.g., Lavvas et al., 2011). Lavvas et al. (2011) modeled a hypothetical heterogeneous process assuming that the adsorption of $C_2H$ onto the surface proceeds at the same rate of the reaction of $C_2H$ with $C_6H_6$. Our experimental results suggest that surface growth by $CH_3$ addition is an important process for the growth of aerosol particles and atmospheric chemistry throughout Titan's middle atmosphere, given the formation of large numbers of $CH_3$ radicals through direct $CH_4$ photolysis at altitudes from 500–900 km and catalytic $CH_4$ dissociation induced by $C_4H_2$ photolysis at altitudes from 200–500 km (Krasnopolsky, 2014).

Figure 11 illustrates a schematic image of physcial processes regarding the formation and growth of Titan's organic aerosols. Our experimental results suggest that aerosol particles are coated with aliphatic and/or olefinic hydrocarbons through heterogeneous reactions of $CH_3$ in the middle atmosphere of Titan, suggesting that the aerosols may form a core-mantle structure duirng the precipitation in the atmosphere. The initial aerosol particles formed in the upper atmosphere would contain N-bearing polyaromatic compounds (Waite et al., 2007); whereas the coated mantle may have polymer-like olefinic structures. To investigate whether the mantle of polymer-like structure affects the optical constants of aerosol particles, further microphysical studies are needed. A quantitative evaluation of the relative importance of heterogeneous reactions to gas-phase nucleation reactions for total aerosol production requires an introduction of the reaction probability of $CH_3$ addition into a 1-D photochemical model, which will be a future work of the present study.

We investigate the influence of $CH_3$ addition on the carbon budget on Titan's surface by estimating the amount of carbon incorporated into aerosols and reduction of total $C_2H_6$ production. Using the above equations (1) and (2), we estimate the rate of $CH_3$ addition onto aerosols' surfaces at altitude from 150–250 km, where thick haze layers exist. Based on the atmospheric temperature ($T \sim 180$ K) and the number density of $CH_3$ radicals ($10^6$ cm$^{-3}$) at 150–250 km (Krasnopolsky, 2009), collision flux of $CH_3$ radicals to the surfaces of aerosol particles is estimated to be $1.2 \times 10^{10}$ cm$^{-2}$ s$^{-1}$. The surface area of aerosol particles at ~200 km has been estimated as on the order of ~$10^{-7}$–$10^{-6}$ cm$^2$ cm$^-$ [3] (Toon et al., 1992; Rannou et al., 2003; Lavvas et al., 2010). Assuming the surface area of aerosol particles in the atmosphere at 200 km ($10^{-7}$–$10^{-6}$ cm$^2$ cm$^{-3}$) together with the reaction probability of ~$10^{-2}$ for $CH_3$ addition, the addition rate of $CH_3$ is estimated as 10–100 cm$^{-3}$ s$^{-1}$. The column rate of $CH_3$ addition onto aerosol particles then becomes $10^8$–$10^9$ cm$^{-2}$ s$^{-1}$, or $10^{-15}$–$10^{-14}$ g cm$^{-2}$ s$^{-1}$ of carbon, assuming the addition reaction proceeds homogeneously from 150 km to 250 km in altitude. The estimated column rate of $CH_3$ addition is comparable to the proposed production rate of aerosols in Titan's atmosphere, i.e., $(0.5$–$2) \times 10^{-14}$ g cm$^{-2}$ s$^{-1}$ (McKay et al., 2011), suggesting that the $CH_3$ addition would be one of the major process in the growth of organic aerosols in Titan's atmosphere. The column rate of $CH_3$ addition, $10^8$–$10^9$ cm$^{-2}$ s$^{-1}$, is also comparable with the $C_2H_6$ column loss rate due to condensation at the tropopause ($5.2 \times 10^8$ cm$^{-2}$ s$^{-1}$, Krasnopolsky, 2014).

In Titan's atmosphere, $CH_3$ radicals are converted mainly into $C_2H_6$ through the following three-body reaction: $CH_3 + CH_3 + M \rightarrow C_2H_6 + M$ (Krasnopolsky, 2009). Accordingly, the net column production rate of $C_2H_6$ would be reduced at least ~10% considering the consumption of $CH_3$ radicals on aerosol surfaces. To further assess the role of heterogeneous reactions of $CH_3$ addition for the carbon budget on Titan more quantitatively, an incorporation of this reaction into a one-dimensional photochemical model would be required. Nevertheless, the present study suggests the possibility that $CH_3$ addition remarkably affects the fate of dissociated $CH_4$ in the middle atmosphere and the carbon budget on Titan.

*4.2.2. Organic aerosol production on early Earth*

Indirect greenhouse effect due to UV shielding of $NH_3$ by organic aerosol layers has been proposed as one of the possible solutions for the faint young Sun paradox (Sagan and Chyba, 1997; Pavlov et al., 2001; Trainer et al., 2006; DeWitt et al., 2009; Wolf and Toon, 2010). However, the production rate and size of organic aerosols on early Earth remains highly uncertain, which results in significant uncertainties in calculating surface temperatures by radiative transfer models.

Wolf and Toon (2010) pointed out the importance of size and shape of organic aerosols for the indirect greenhouse effect. They estimate the optical depth due to fractal aggregates of organic aerosols for various number densities (or production rate) and size of monomers. They show that organic aerosols would have provided sufficient indirect greenhouse effect, assuming an annual production of $10^{14}$ g year$^{-1}$ and a monomer radius of 50 nm based on the size of Titan's aerosols (Tomasko et al., 2008; Lavvas et al., 2010). The number density of aerosol monomers would be determined by gas-phase nucleation of aerosols in the upper atmosphere; whereas, the size of monomers may be affected by heterogeneous growth during the precipitation in the atmosphere. Thus, the size of monomers could vary between early Earth and Titan if the efficiency of heterogeneous growth during the precipitation is different.

Trainer et al. (2006) estimated the production rate of organic aerosols by FUV irradiation using a scaling law, based on Titan's aerosol production rate and their experimental results of $CH_4/CO_2$ dependency of aerosol formation rate. The scaling law of Trainer et al. (2006) takes into account the dependence of aerosol production rate on UV flux and $CH_4/CO_2$ ratio as follows,

$$F = \beta \cdot F_\mathrm{T} \cdot \left(\frac{I}{I_\mathrm{T}}\right)^m \cdot \left(\frac{\chi}{\chi_\mathrm{T}}\right) \qquad (3)$$

where $F$ and $F_T$ are the production rates of organic aerosols on early Earth and Titan, respectively; $I$ and $I_T$ are the solar UV fluxes received by early Earth and Titan, respectively; $\chi$ and $\chi_T$ are the atmospheric mixing ratios of $CH_4$ on early Earth and Titan, respectively; $m$ is an exponent corresponding to the dependence on actinic flux (Trainer et al., 2006). The term $\beta$ is a scaling factor to account for the dependence of organic aerosol production rate on $CH_4/CO_2$ ratio, where $\beta = 1$ for a pure $CH_4$ atmosphere (Trainer et al., 2006).

Our experimental results show that the growth rate of organic film at $CH_4/CO_2 = 1$ is ~$3 \times 10^{-2}$ times that for pure $CH_4$ gas (i.e., $\beta = 0.03$) (Fig. 4). In addition, for the first-order photolysis reaction including $CH_3$ production, $m$ becomes unity (Trainer et al., 2006). Using $\beta = 0.03$ and $m = 1$, together with $I/I_T = 230$ and $\chi/\chi_T = 0.063$ (Trainer et al. 2006, and references therein), the growth rate on early Earth would be 0.4 times that on Titan. Given the shorter timescales for aerosol precipitation in the atmosphere of early Earth (a few years, Wolf and Toon, 2010) compared with that of Titan (a hundred years, Toon et al. 1992), the amount of aerosols grown by $CH_3$ addition on early Earth could have been considerably lower than that on Titan, resulting in the formation of smaller-sized monomers. This could affect early Earth's climate through changing effective optical depth due to aerosol layers (Wolf and Toon, 2010).

In the above discussion, we assumed that the parameter $\beta$ which describes relative aerosol production rate is fully controlled by heterogeneous reaction. However, aerosol nucleation is also an important factor to control production rate in real atmospheres, which we did not investigate this process in our experiments. Further theoretical and experimental study is needed to assess whether $CH_3$ addition can significantly influence the total production rate of organic aerosols.

**5. Conclusions**

We conducted laboratory experiments simulating organic photochemistry by solar FUV irradiation in a $CH_4$-containing atmosphere. Using a $H_2$-He lamp as a FUV source, which has a UV spectrum similar to that of solar FUV, we performed laboratory experiments to investigate the growth rate of organic film and the gaseous products from $CH_4$-$CO_2$ gas mixtures as a function of $CH_4/CO_2$ ratio. To interpret the mechanism to form organic films in our experiments, we also performed one-box photochemical model calculations to simulate photochemistry occurred during the FUV irradiation. Based on the experimental results, we can summarize our experimental and calculation results as follows:

- The growth rate of the organic film exhibits a steep decrease with the $CH_4/CO_2$ ratio being less than unity. The growth rate does not have a maximum at around $CH_4/CO_2$ =1, in contrast to the previous results given by Trainer et al. (2006).
- Results of the photochemical model show that the polymerization reactions involving aromatic hydrocarbons have a similar $CH_4/CO_2$ dependence with the observed growth rates of the organic films. However, the calculated growth rate of the produced aerosol particles by the reactions of aromatic hydrocarbons cannot account for the measured growth rate of the organic films quantitatively.
- Calculation results of $CH_4/CO_2$ dependency of polymerization rates of polyynes show a poor agreement with the experimental results of the growth rate of the organic film, suggesting that polymerization of polyynes is also not the major process for formation of the organic films.

- Heterogeneous reaction of $CH_3$ addition onto the organic film can explain the measured growth rate of the organic films at the reaction probability of ~$10^{-2}$ for $CH_4/CO_2 > 3$.
- These results suggest that the surface growth proceeds mainly through the heterogeneous reactions of $CH_3$ addition in our experiments. At low $CH_4/CO_2$, etching of the organic film by O atom formed by $CO_2$ photolysis could be a competitive process to reduce or inhibit the growth of the organic films.

Based on these results, we discuss the roles of heterogeneous reactions in the atmosphere of Titan and early Earth for the aerosol production and consequences on surface environments. Although nucleation process is not considered in the present study, we suggest that the heterogeneous reaction of $CH_3$ addition would reduce the production and precipitation of $C_2H_6$ in Titan's atmosphere. Using the scaling law proposed by Trainer et al. (2006), we suggest that aerosol growth in early Earth could have been ineffective compared with that of current Titan, leading to the formation of smaller-sized monomers of aerosols in early Earth's atmosphere.

**6. Acknowledgements**

The authors acknowledge H. Kuwahara for his help to obtain the data of fragmentation pattern for mass spectrum analyses; H. Imanaka for constructive comments at the start of this work; and HORIBA, Co. LTD., for their technical supports in ellipsometry. P. Hong acknowledges support from TeNQ/Tokyo-dome. This work was supported by KAKENHI (Grant Numbers 11J06087, 25120006, and 26707024) from Japan Society for Promotion of Science (JSPS), KAKENHI (Grant Number 23103003 and JP17H06456) from the Ministry of Education, Culture, Sports, Science and

**Supplementary materials**

Supplementary material associated with this article can be found, in the online version, at doi: 10.1016/j.icarus.2018.02.019.

Table 1. Hydrocarbon reactions previously assumed to produce aerosol particles (i.e., polymerization reactions described in the main text). See Appendix for the reaction rates.

| Reaction | Previous models |
| --- | --- |
| R205  $C + C_4H_6 \rightarrow C_3H_3 + C_2H_3$ | Lavvas et al. (2008a,b) |
| R219  $CH + C_4H_8 \rightarrow C_5H_8 + H$ | Lavvas et al. (2008a,b), Wilson and Atreya (2004) |
| R310  $C_2 + C_6H_6 \rightarrow$ Products | Hébrard et al. (2006) |
| R314  $C_2H + C_2H_2 \rightarrow C_4H_2 + H$ | Pavlov et al. (2001) |
| R322  $C_2H + CH_2CCH_2 \rightarrow C_5H_4$ | Pavlov et al. (2001) |
| R337  $C_2H + C_6H_6 \rightarrow$ Products | Wilson and Atreya (2004), Hébrard et al. (2006) |
| R338  $C_2H + C_8H_2 \rightarrow$ Products + H | Yung et al. (1984), Toublanc et al. (1995), Wilson and Atreya (2004), Hébrard et al. (2006), Krasnopolsky (2009, 2010) |
| R468  $C_4H + C_6H_2 \rightarrow$ Products + H | Yung et al. (1984), Toublanc et al. (1995), Wilson and Atreya (2004), Hébrard et al. (2006), Lavvas et al. (2008a,b), Krasnopolsky (2009, 2010) |
| R469  $C_4H + C_8H_2 \rightarrow$ Products + H | Yung et al. (1984), Toublanc et al. (1995), Wilson and Atreya (2004), Hébrard et al. (2006), Krasnopolsky (2009, 2010) |
| R511  $C_6H + C_4H_2 \rightarrow$ Products+ H | Yung et al. (1984), Toublanc et al. (1995), Wilson and Atreya (2004), Hébrard et al. (2006), Lavvas et al. (2008a,b) |

| R512 | $C_6H + C_6H_2 \rightarrow$ Products + H | Yung et al. (1984), Toublanc et al. (1995), Wilson and Atreya (2004), Hébrard et al. (2006), Lavvas et al. (2008a,b), Krasnopolsky (2009, 2010) |
|---|---|---|
| R513 | $C_6H + C_8H_2 \rightarrow$ Products + H | Yung et al. (1984), Toublanc et al. (1995), Wilson and Atreya (2004), Hébrard et al. (2006), Krasnopolsky (2009, 2010) |
| R515 | $C_6H_5 + C_2H_2 \rightarrow$ Products + H | Wilson and Atreya (2004), Hébrard et al. (2006), Lavvas et al. (2008a,b), Krasnopolsky (2009, 2010) |
| R516 | $C_6H_5 + C_2H_2 \rightarrow$ Products | Wilson and Atreya (2004), Hébrard et al. (2006), Lavvas et al. (2008a,b), Krasnopolsky (2009, 2010) |
| R517 | $C_6H_5 + C_6H_6 \rightarrow$ Products + H | Wilson and Atreya (2004), Hébrard et al. (2006), Krasnopolsky (2009, 2010) |
| R661 | $O(^3P) + C_3H_3 \rightarrow$ Products + H | Wilson and Atreya (2004), Hébrard et al. (2006) |
| R662 | $O(^3P) + C_3H_5 \rightarrow$ Products + H | Hébrard et al. (2006) |
| R783 | $CH_2OH + C_2H_4 \rightarrow$ Products | Hébrard et al. (2006) |
| R814 | $CH_3CO + C_2H_3 \rightarrow$ Products + $CH_3$ | Hébrard et al. (2006) |

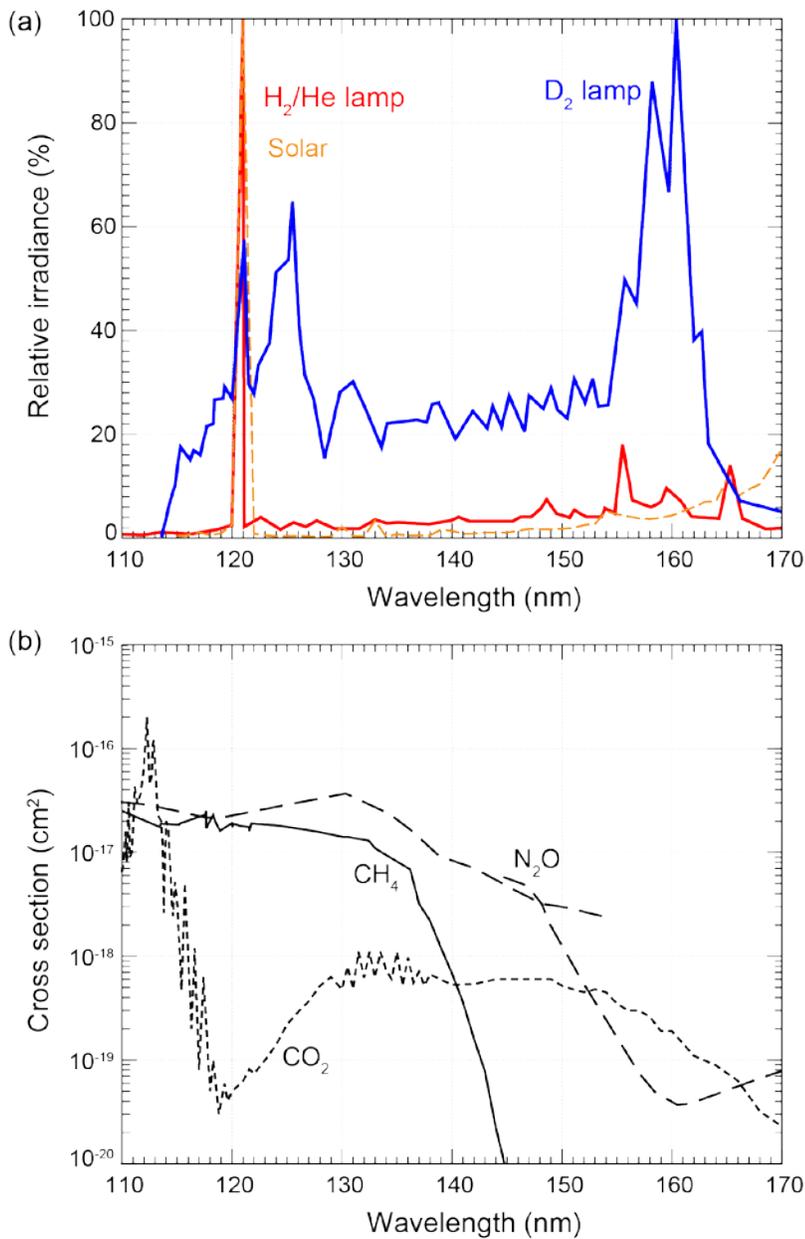

Figure 1. (a) Far ultraviolet spectra of hydrogen-helium lamp obtained by Kuroiwa et al., 1992) (solid red line), deuterium lamp (Hamamatsu Photonics Co., L1835) (solid blue line) and solar ultraviolet flux (Mount and Rottman, 1983) (yellow dashed line). (b) Absorption cross sections of $CH_4$ (solid line), $CO_2$ (broken line) compiled by the Southwest Research Institute[2], and $N_2O$ (Chan et al., 1994; Nicolet and Peetermans, 1972) (dashed line).

---

[2] http://phidrates.space.swri.edu/

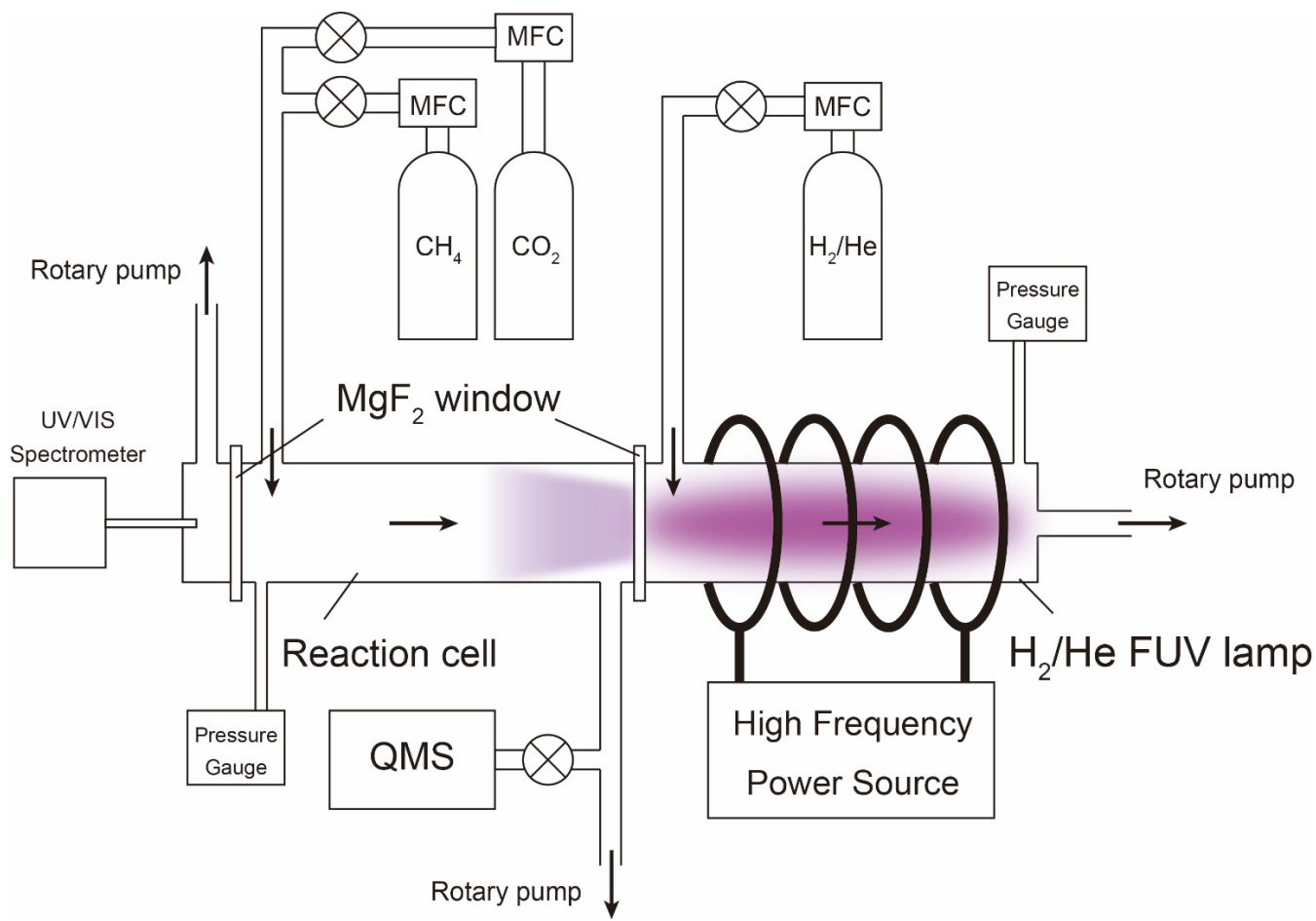

Figure 2. A schematic diagram of the experimental apparatus for organic aerosol production. QMS and MFC stand for Quadrupole Mass Spectrometer and Mass Flow Controller, respectively. Black arrows represent the directions of gas flows of reactant gas mixtures of $CH_4$ and $CO_2$ and pre-mixed $H_2$-He gas.

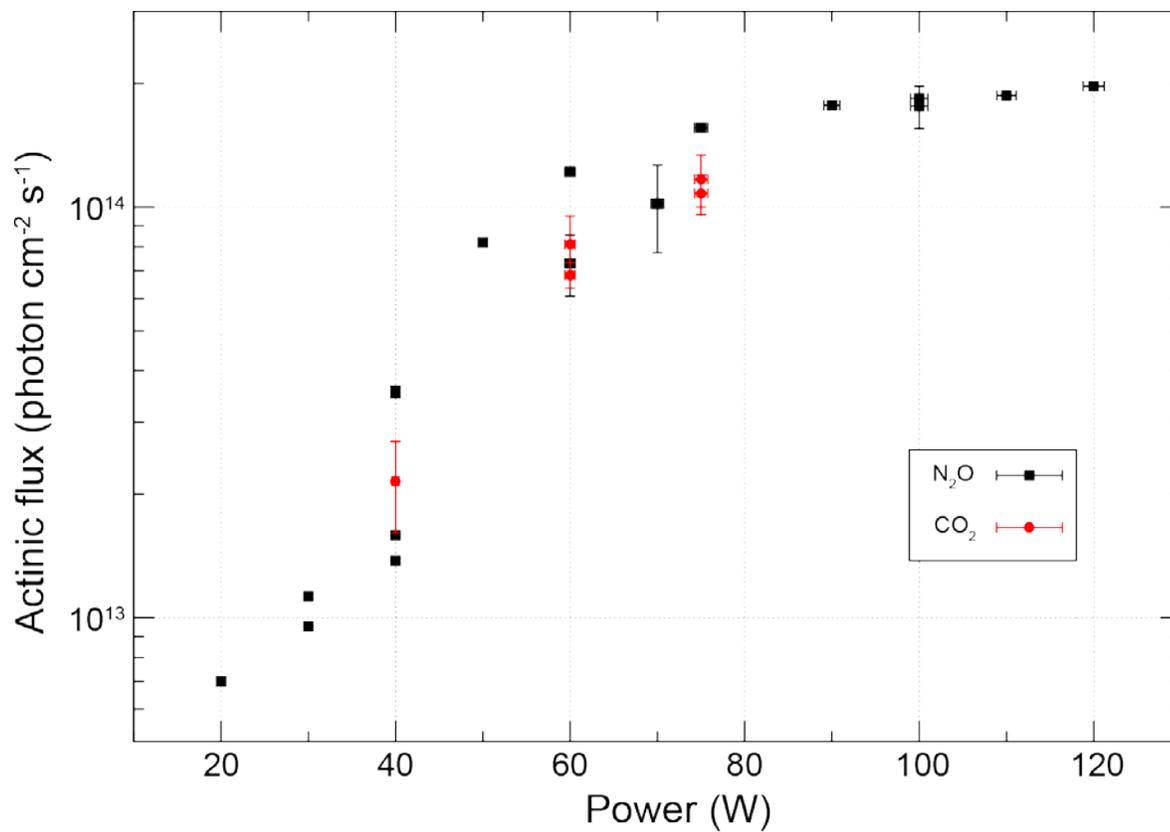

Figure 3. Results of actinometry for the H$_2$-He lamp. Black squares and red circles show the results of N$_2$O and CO$_2$ actinometry, respectively.

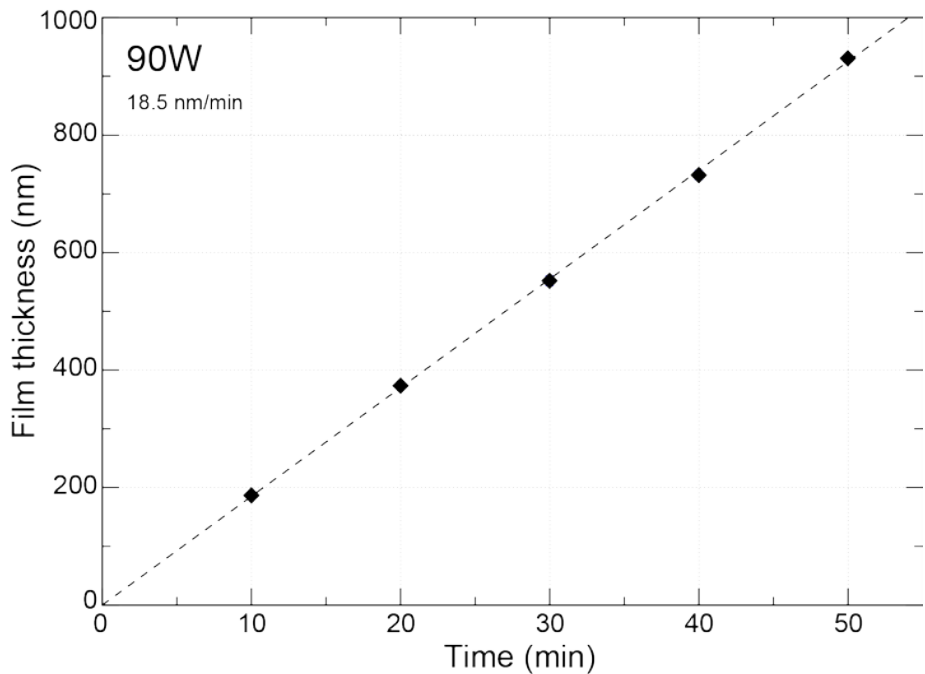

Figure 4. Film thickness as a function of irradiation time for pure $CH_4$ gas experiment with RF power of 90 W. Broken line shows the best-fit linear function whose slope, 18.5 nm/min for this case, is the growth rate of organic film. The measurement error of film thickness by ellipsometry is ±1 nm for this case.

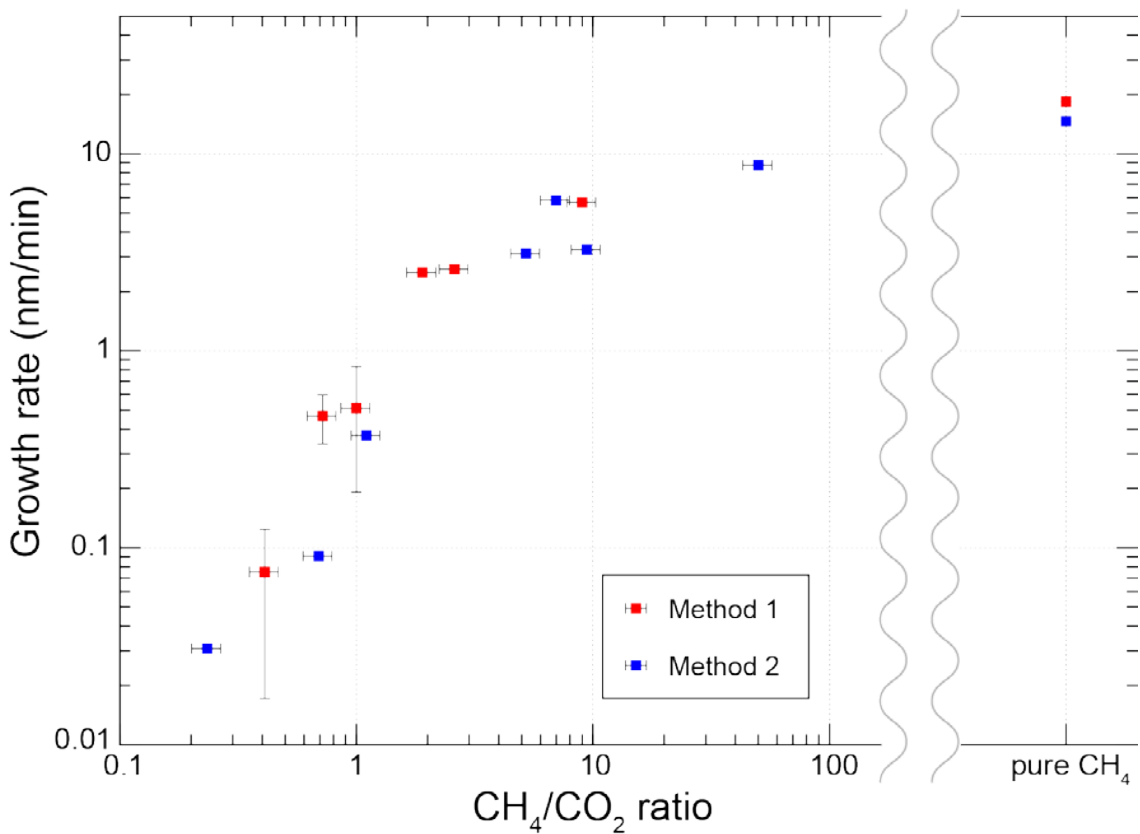

Figure 5. Growth rate of the organic film as a function of initial $CH_4/CO_2$ ratio. Red and blue squares represent the results measured with Methods 1 and 2, respectively (see the main text). The errors of $CH_4/CO_2$ ratio were caused by the measurement error of influx through mass flow controllers. The relative error of $CH_4/CO_2$ ratio was estimated to be 14% based on propagation of error, since each flux measurement for the gases contains 10% of error.

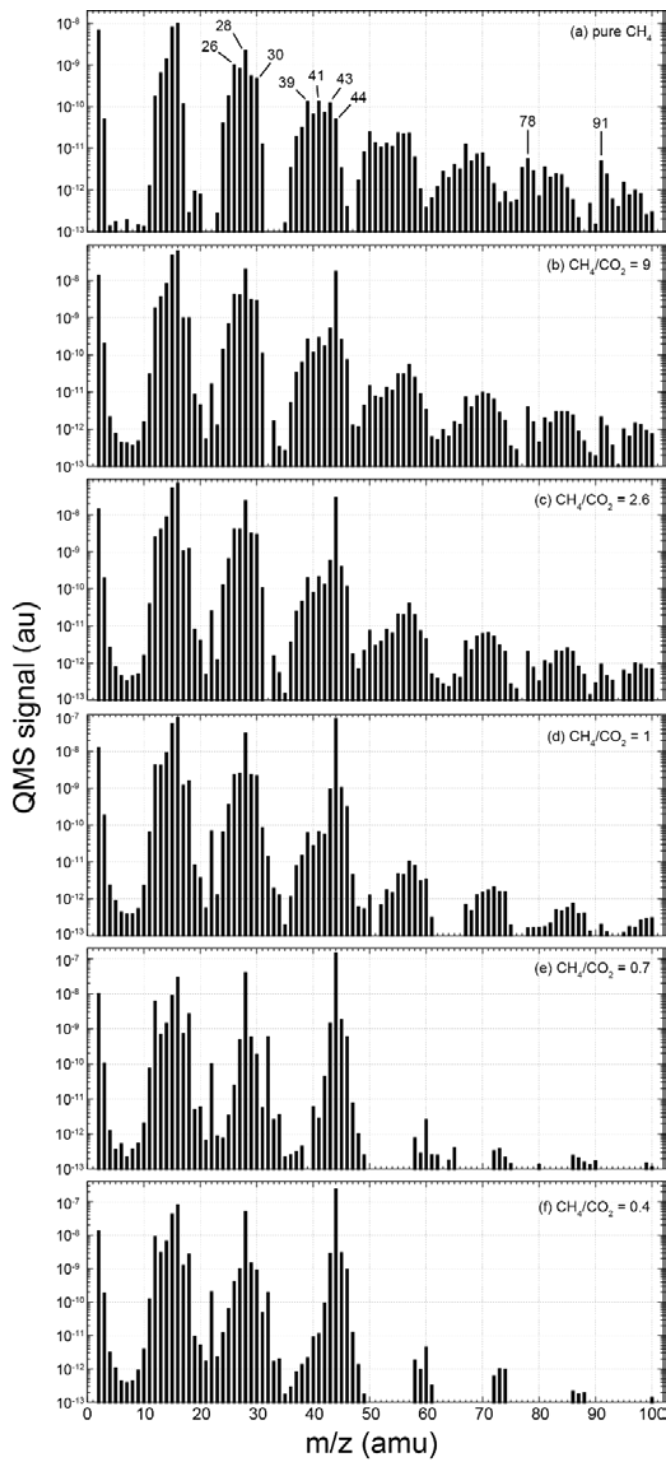

Figure 6. Mass spectra of gas mixtures after the photochemical reactions. Initial reactant gases are (a) pure CH$_4$, (b) CH$_4$/CO$_2$ = 9, (c) CH$_4$/CO$_2$ = 2.6, (d) CH$_4$/CO$_2$ = 1, (e) CH$_4$/CO$_2$ = 0.7, (f) CH$_4$/CO$_2$ = 0.4. Background signals are subtracted from the mass spectra (see the Supplemental Material).

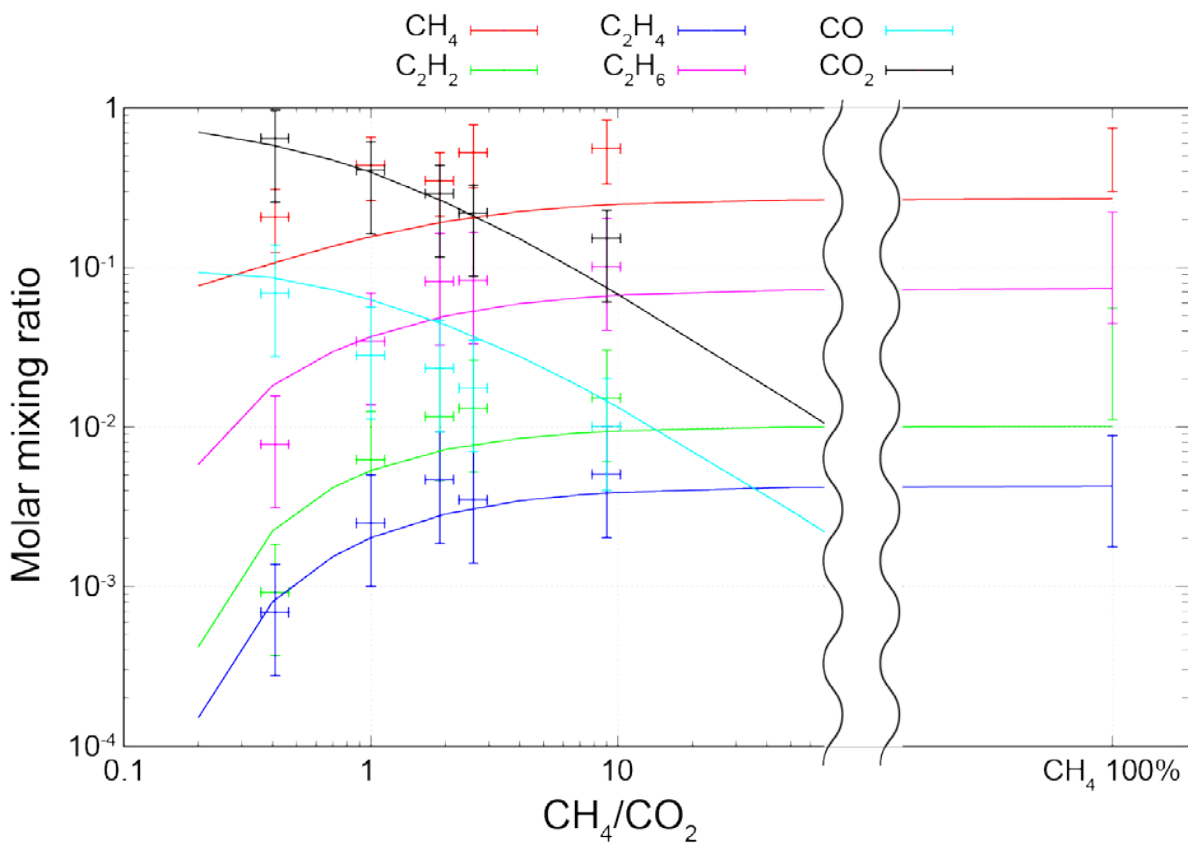

Figure 7. Molar mixing ratios of major gas species retrieved by the deconvolution analysis of mass spectra (crosses) and calculated with the one-box photochemical model (lines). The abundances calculated by the photochemical model represent steady-state abundances. The errors of mixing ratios were resulted mainly from the deconvolution analysis of the mass spectra. The errors of $CH_4/CO_2$ ratio were caused by the measurement error of influx through mass flow controllers. The relative error of $CH_4/CO_2$ ratio was estimated to be 14% based on propagation of error, since each flux measurement for the gases contains 10% of error.

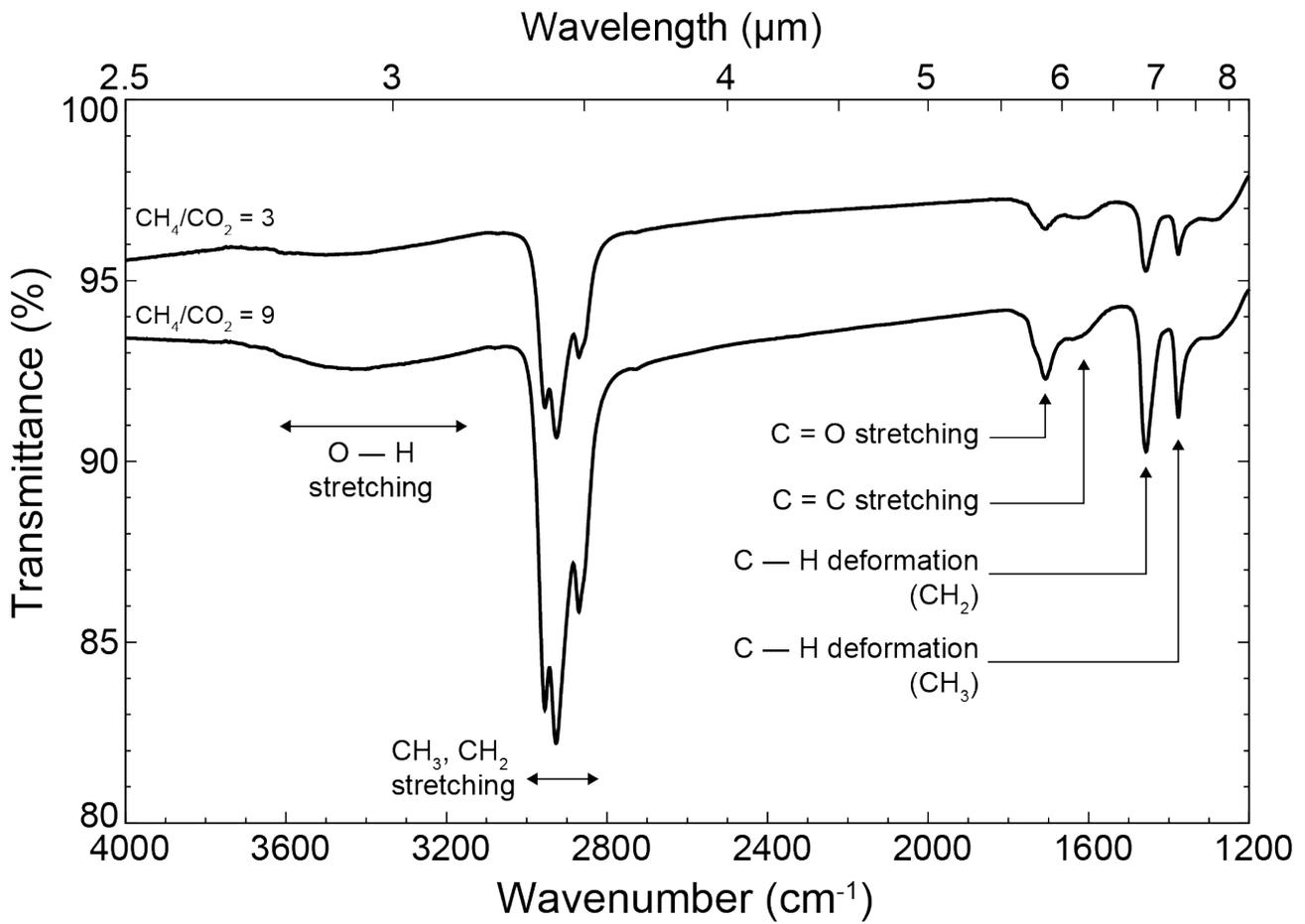

Figure 8. Typical infrared transmittance spectra of organic films deposited on MgF$_2$ window formed from different reactant gas mixtures. The transmittances have been offset.

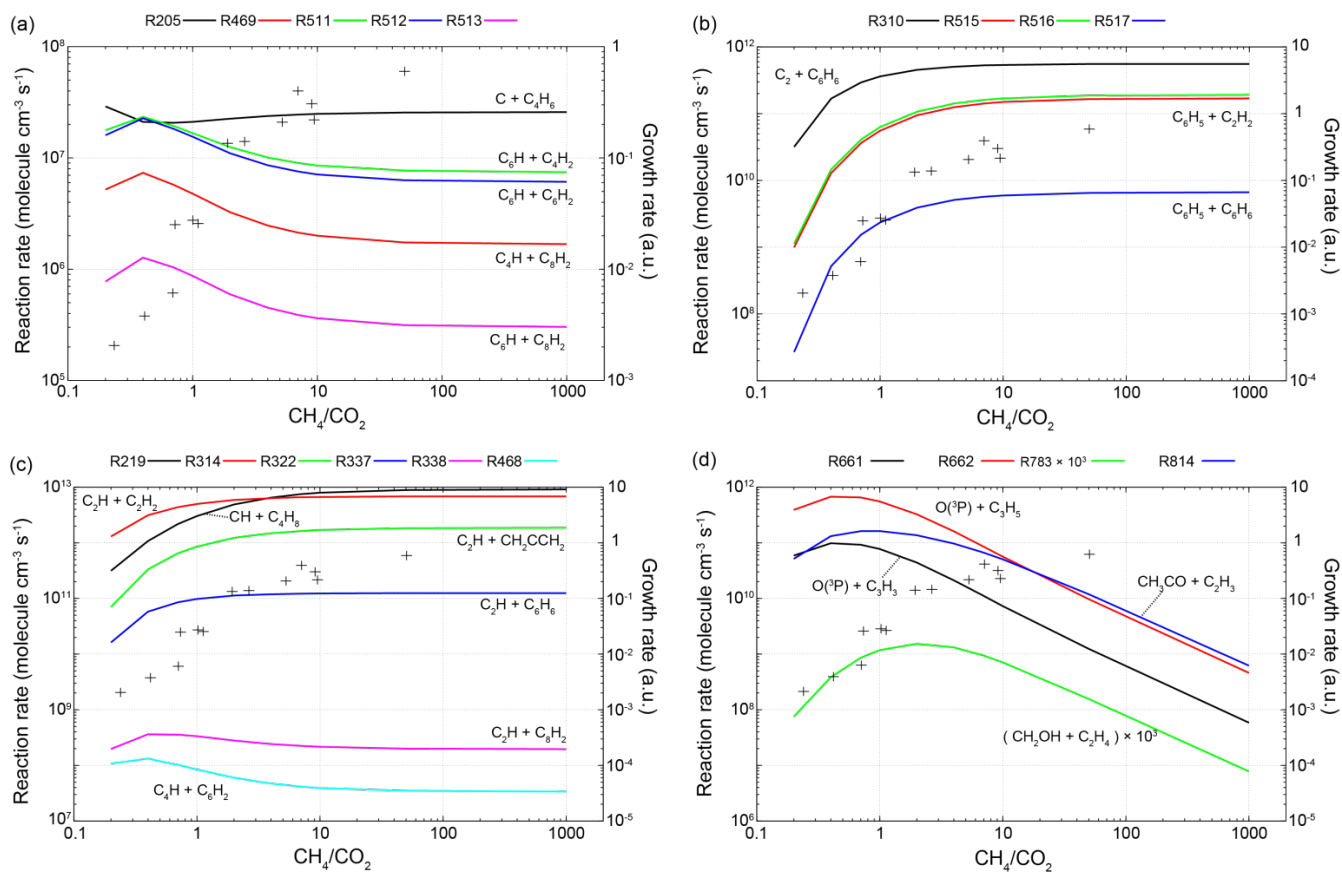

Figure 9. Steady-state rates of polymerization reactions, involving (a) C, $C_4H_2$, $C_6H_2$, $C_8H_2$, (b) $C_6H_5$, $C_6H_6$, (c) CH, $C_2H$, and $C_4H$, (d) oxygen species, compared with the growth rates of the organic films (crosses). The reaction numbers are same as those in Table 1 and Appendix. The growth rates are normalized by the result obtained for pure $CH_4$ gas.

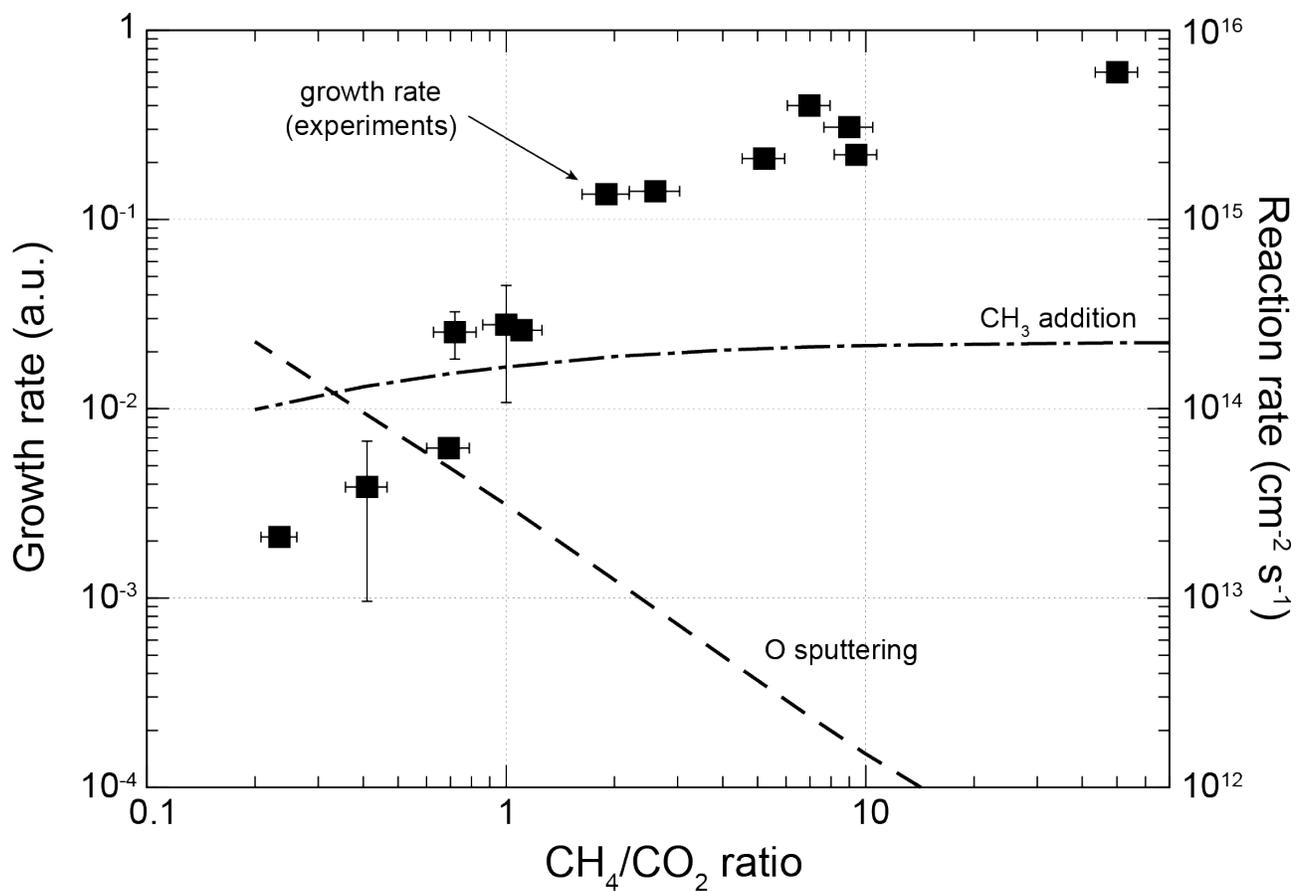

Figure 10. Comparison of heterogeneous reaction rates (lines) and the growth rate of the organic films (squares) as functions of $CH_4/CO_2$ ratio for our experimental condition.

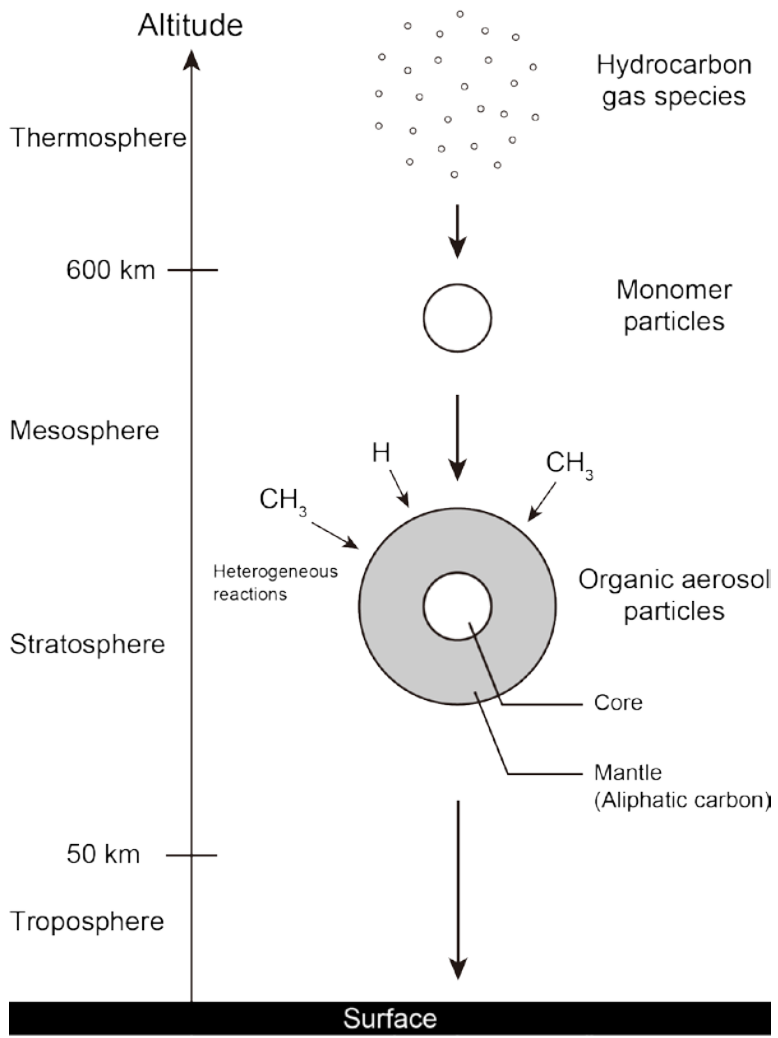

Figure 11. A schematic of the physical processes of Titan's organic aerosols.